\documentclass[aps,prd,nofootinbib,amsmath,amssymb,showpacs,superscriptaddress,twocolumn,10pt]{revtex4}
\usepackage{txfonts}
\usepackage{graphicx}
\usepackage{dcolumn}
\usepackage{bm}
\usepackage{amssymb}
\usepackage{latexsym}
\usepackage{booktabs}
\usepackage[colorlinks=true, linkcolor=red, citecolor=blue]{hyperref}

\newcommand{\be}{\begin{equation}}
\newcommand{\ee}{\end{equation}}
\newcommand{\bq}{\begin{eqnarray}}
\newcommand{\eq}{\end{eqnarray}}

\begin{document}

\title{A global fit study on the new agegraphic dark energy model}

\author{Jing-Fei Zhang}
\affiliation{Department of Physics, College of Sciences,
Northeastern University, Shenyang 110004, China}
\author{Yun-He Li}
\affiliation{Department of Physics, College of Sciences,
Northeastern University, Shenyang 110004, China}
\author{Xin Zhang\footnote{Corresponding author}}
\email{zhangxin@mail.neu.edu.cn} \affiliation{Department of Physics,
College of Sciences, Northeastern University, Shenyang 110004,
China} \affiliation{Center for High Energy Physics, Peking
University, Beijing 100080, China}

\begin{abstract}
We perform a global fit study on the new agegraphic dark energy
(NADE) model in a non-flat universe by using the MCMC method with
the full CMB power spectra data from the WMAP 7-yr observations, the
SNIa data from Union2.1 sample, BAO data from SDSS DR7 and WiggleZ
Dark Energy Survey, and the latest measurements of $H_0$ from HST.
We find that the value of $\Omega_{k0}$ is greater than $0$ at least
at the 3$\sigma$ confidence levels (CLs), which implies that the
NADE model distinctly favors an open universe. Besides, our results
show that the value of the key parameter of NADE model,
$n=2.673^{+0.053+0.127+0.199}_{-0.077-0.151-0.222}$, at the
1--3$\sigma$ CLs, where its best-fit value is significantly smaller
than those obtained in previous works. We find that the reason
leading to such a change comes from the different SNIa samples used.
Our further test indicates that there is a distinct tension between
the Union2 sample of SNIa and other observations, and the tension
will be relieved once the Union2 sample is replaced by the Union2.1
sample. So, the new constraint result of the NADE model obtained in
this work is more reasonable than before.
\end{abstract}

 \maketitle

\section{Introduction}\label{sec1}

Dark energy has become one of the most important research areas in
cosmology since the cosmic acceleration was discovered by the
observations of type Ia supernovae (SNIa)~\cite{Riess98}. The
cosmological constant, $\Lambda$, introduced by Einstein in 1917, is
a natural candidate for dark energy, which fits the observational
data best to date. However, the model of $\Lambda$ plus cold dark
matter, known as the $\Lambda$CDM model, always suffers from the two
well-known theoretical difficulties, namely, the fine-tuning and
cosmic coincidence problems~\cite{dereview}. Thus, dynamical dark
energy models become popular, because they may alleviate the
theoretical challenges faced by the $\Lambda$CDM model. For a recent
review of dark energy, see Ref.~\cite{Li:2011sd}.

Among the many dynamical dark energy models, a class with feature of
quantum gravity, embodying holographic principle, looks very special
and attractive. Such class of models, usually called ``holographic
dark energy'', has been investigated in detail~\cite{Li:2004rb} and
constrained by the observational data~\cite{holofit}. In this paper,
we focus on the ``new agegraphic dark energy'' (NADE)
model~\cite{Wei:2007ty} in this class, which takes into account the
uncertainty relation of quantum mechanics together with the
gravitational effect in general relativity. The reason that the NADE
model is fairly attractive is also owing to the fact that this model
has the same number of parameters to the $\Lambda$CDM model, less
than other dynamical dark energy models. The NADE model has been
proven to fit the data well. However, it should be noticed that the
global fit analysis on the NADE model has never been done.

The global fit analysis is very important for the study of a
cosmological model. To determine the parameters of dark energy
models, one often has them constrained by using some important
observational data sets, such as SNIa, baryon acoustic oscillation
(BAO) and cosmic microwave background (CMB). For the CMB
information, to save time and power, one often uses the CMB
``distance priors'' ($R$, $l_A$, $z_{\ast}$) whose information is
extracted from the temperature power spectrum of CMB based on a
$\Lambda$CDM scenario. However, it should be noted that using the
CMB ``distance priors'' to constrain other dark energy models will
lead to a circular problem, since the values of the CMB ``distance
priors'' depend on a specific cosmological model, namely, the
$\Lambda$CDM model. Besides, the ``distance priors'' do not capture
the information concerning the growth of structure probed by the
late-time ISW effect which is sensitive to the properties of dark
energy. In the past, the NADE model was constrained only by the CMB
``distance priors'' data. In this study, we shall use the full CMB
data from the 7-yr WMAP observations to explore the parameter space
of the cosmological model involving the NADE. Such an exploration is
believed to complete the study of the NADE model.

In the previous studies, the spatial curvature, $\Omega_{k0}$, is
usually neglected. So, we do not know if a flat universe is still
favored by current data in the NADE model. Moreover, the initial
condition of the NADE model~\cite{Wei:2007ty} is only applicable for
a flat universe containing only dark energy and pressureless matter.
In Ref.~\cite{Li:2012xm}, a new, updated initial condition is
proposed, which is applicable for a non-flat universe with various
components. We shall apply this new initial condition in the present
work. In such an application, for employing the ``CosmoMC'' package,
one may treat the parameter $n$ as a derived parameter and use the
Newton's iteration algorithm to determine its value. In addition, in
a global fit analysis, the cosmological perturbations of dark energy
should also be considered. Thanks to the quintessence-like property
of the NADE, there is no $w=-1$-crossing behavior and so the
corresponding gravity instability is absent. In the holographic dark
energy model, the $w=-1$-crossing happens, so one has to employ the
parameterized post-Friedmann (PPF) approach to treat the gravity
instability of the perturbations in a global fit analysis; for
details see Ref.~\cite{Wang:2012uf}. For the NADE model, however, we
have no need to employ the PPF approach.

Our paper is organized as follows. In Sec.~\ref{sec2}, we derive the
background and perturbation evolution equations for the NADE model
in a non-flat universe. In Sec.~\ref{sec3}, we introduce the fit
method and the observational data, and then give the global fit
results. Concluding remarks are given in Sec.~\ref{sec4}. In this
work, we assume today's scale factor $a_0=1$, and so the redshift
$z$ satisfies $z = a^{-1}-1$; the subscript ``0'' always indicates
the present value of the corresponding quantity, and the units with
$c=\hbar=1$ are used.

\section{Brief description of the NADE model}\label{sec2}

The NADE model is constructed in light of the K\'{a}rolyh\'{a}zy
relation and corresponding energy fluctuations of space-time. The
energy density of NADE, $\rho_{de}$, has the form~\cite{Wei:2007ty}
 \be\label{eq:rhoq}
   \rho_{de}=\frac{3n^2M_{Pl}^2}{\eta^2},
 \ee
where $n$ is a numerical parameter, $M_{Pl}$ is the reduced Planck
mass, and $\eta$ is the conformal age of the universe,
 \be\label{eq:eta}
   \eta\equiv\int_0^t\frac{d\tilde{t}}{a}=\int_0^a\frac{d\tilde{a}}{H\tilde{a}^2},
 \ee
where $a$ is the scale factor and $H\equiv\dot{a}/a$ is the Hubble
parameter. Here a dot denotes the derivative with respect to the
cosmic time $t$.

The background evolution of a non-flat universe is described by the
Friedmann equation,
 \be\label{eq:Friedmann}
  \sum\Omega_i=1,
 \ee
where $\Omega_i$ is defined as the ratio of the energy density
$\rho_i$ to the critical energy density $\rho_c\equiv{3M_{Pl}^2H^2}$
with $i=de$, $dm$, $b$, $r$ and $k$ which denotes the dark energy,
dark matter, baryon, radiation and spatial curvature, respectively.
Note that here $\rho_k\equiv-3M_{Pl}^2k/{a^2}$.

The energy conservation equations for the various components take
the form \be\label{eq:conservation}
  \dot\rho_i+3H(1+w_i)\rho_i=0,
\ee where $w_i$ is the equation-of-state parameter (EOS) of a
specific component denoted by subscript $i$, such as $w_{dm}=w_b=0$,
$w_r=1/3$ and $w_k=-1/3$. For the EOS of NADE, one can obtain
 \be\label{eq:EoS}
  w_{de}=-1+\frac{2(1+z)}{3n}\sqrt{\Omega_{de}},
 \ee
from Eqs.~(\ref{eq:rhoq}), (\ref{eq:eta}) and
(\ref{eq:conservation}). Furthermore, using
Eqs.~(\ref{eq:rhoq})--(\ref{eq:conservation}), one can also easily
get the equation of motion of $\Omega_{de}$, a differential
equation,
 \be\label{eq:OqzEoM}
  \frac{d\Omega_{de}}{dz}=\frac{-\Omega_{de}}{1+z}\left(1-\Omega_{de}\right)\left[3+G(z)-\frac{2(1+z)}{n}\sqrt{\Omega_{de}}\right],
 \ee
where
$G(z)=\frac{\Omega_{r0}(1+z)^2-\Omega_{k0}}{\Omega_{m0}(1+z)+\Omega_{r0}(1+z)^2+\Omega_{k0}}$
with $\Omega_{m0}=\Omega_{dm0}+\Omega_{b0}$.

For solving the differential equation~(\ref{eq:OqzEoM}), we adopt
the initial condition,
$\Omega_{de}(z_{ini})=\frac{n^2(1+z_{ini})^{-2}}{4}\left(1+\sqrt{F(z_{ini})}\right)^2$
at $z_{ini}=2000$ with
$F(z)=\frac{\Omega_{r0}(1+z)}{\Omega_{m0}+\Omega_{r0}(1+z)}$, given
in Ref.~\cite{Li:2012xm}, which helps reduce one free parameter for
the NADE model. This initial condition is an updated version of the
original one, $\Omega_{de}(z_{ini})=n^2(1+z_{ini})^{-2}/4$ at
$z_{ini}=2000$, given in Ref.~\cite{Wei:2007xu}. The only difference
between them is that the new one is valid when the radiation is
considered as a non-ignorable component. Note that the initial
condition is also applicable in a non-flat universe, since the
spatial curvature $\Omega_k$ is much smaller than $\Omega_m$ or
$\Omega_r$ at $z = 2000$. For more details about the derivation and
application of the new initial condition, see Ref.~\cite{Li:2012xm}.

In our work, for simplicity, we fix
$\Omega_{r0}=2.469\times10^{-5}h^{-2}(1 + 0.2271N_{\rm eff})$
according to the WMAP 7-yr observations~\cite{WMAP7}, where $N_{\rm
eff}=3.04$ is the standard value of the effective number of neutrino
species and $h$ is the Hubble constant $H_0$ in units of 100
km/s/Mpc.

It should be noted that the parameters $n$, $\Omega_{dm0}$,
$\Omega_{b0}$ and $\Omega_{k0}$ are not independent. One of them can
be derived from the others. For example, we can treat the parameter
$n$ as the derived parameter. In this case, once the values of
$\Omega_{dm0}$, $\Omega_{b0}$ and $\Omega_{k0}$ are given, one can
impose an initial value of $n$ and then obtain the true value of $n$
by using the Newton iteration algorithm,
$n_{k+1}=n_k-f(n_k)/f'(n_k)$, where
$f(n_k)=1-\Omega_{dm0}-\Omega_{b0}-\Omega_{r0}-\Omega_{k0}-\Omega_{de}(0)|_{n_k}$
with $\Omega_{de}(0)|_{n_k}$ the $k$th order numerical solution (at
$z=0$) of Eq.~(\ref{eq:OqzEoM}) with the new initial condition, and
$f'(n_k)$ is the numerical derivative of $f(n)$ at the point
$n=n_k$. Of course, one can also treat $\Omega_{dm0}$ or
$\Omega_{b0}$ or $\Omega_{k0}$ as a derived parameter and get its
value by using a similar method.

For our global fit analysis, it is convenient to modify the
``CosmoMC'' package~\cite{Lewis:2002ah} if $n$ is set as a derived
parameter. Thus, we will use the Newton method mentioned above,
which is proved to be convergent, to get the true value of $n$. We
choose the initial value $n_0=2.7$ and set a termination condition
for the iteration, such as $|n_{k+1}-n_k|\leq{\epsilon}$ with
$\epsilon$ a small quantity. Once Eq.~(\ref{eq:OqzEoM}) is solved,
the Hubble expansion rate can be obtained by \be\label{eq:hubble}
H(z)=H_0\left[\frac{\Omega_{m0}(1+z)^3+\Omega_{r0}(1+z)^4+\Omega_{k0}(1+z)^2}{1-\Omega_{de}(z)}\right]^{1/2}.
\ee

With the background evolution of the NADE model completed, we now
consider how to handle the perturbation of NADE. Since the EOS of
NADE is always greater than $-1$, we can easily handle its
perturbation. We take NADE as a perfect fluid with the EOS given by
Eq.~(\ref{eq:EoS}). In the synchronous gauge, the conservation of
energy-momentum tensor $T^{\mu}_{\nu;\mu}=0$ gives the perturbation
equations of density contrast and velocity divergence for the NADE
in the Fourier space,
\begin{eqnarray}
 \delta^\prime&=&-(1+w_{de})(\theta+\frac{h^\prime}{2})-3aH(c^{2}_{s}-w_{de})\delta,\\
 \theta^\prime&=&-aH(1-3c^{2}_{s})\theta+\frac{c^{2}_{s}}{1+w_{de}}k^{2}\delta-k^{2}\sigma.
\end{eqnarray}
Here $\prime\equiv d/d\eta$, $H$ and $w_{de}$ are given by
Eqs.~(\ref{eq:hubble}) and (\ref{eq:EoS}). Other notation is
consistent with the work of Ma and Bertschinger~\cite{Ma:1995ey}.
For the gauge ready formalism about the perturbation theory, see
Ref.~\cite{Hwang:2001qk}. For the NADE, we assume the shear
perturbation $\sigma=0$. In our calculations, the adiabatic initial
conditions will be taken.

Finally, we feel that it may be necessary to make some additional
comments on the treatment of dark energy perturbations in the NADE
model. Actually, it is fairly difficult to calculate the
cosmological perturbations of dark energy in the holographic-type
dark energy models, because there is some non-local effect that
makes the perturbation mechanism extremely obscure and hard to
handle. This is the reason why for a long time no one addressed the
issue of dark energy perturbations in a complete manner in the
holographic-type models. Be that as it may, Li, Lin, and
Wang~\cite{Li:2008zq} made a calculation for the perturbations in
the holographic dark energy model, in which some approximations are
used in dealing with the non-local integrals in the equations
governing the evolution of metric perturbations, and they proved
that the perturbations are stable for both super-horizon and
sub-horizon scales. But a complete analysis of the cosmological
perturbations in the holographic dark energy model is still absent.
Thus, in the past, to avoid the inclusion of dark energy
perturbation problem, one only used the CMB ``distance priors''
data, instead of the full CMB data, to constrain the holographic and
agegraphic dark energy models.

However, recently, it was realized that, in order to make progress,
one should first ignore the non-local effect in the holographic-type
models and directly calculate dark energy perturbations in these
models as if they are usual perfect fluids. Alone this line, global
fit analyses on the holographic dark energy model were
performed~\cite{Wang:2012uf,Xu:2012aw}. It should be stressed,
however, that such a treatment is only an expedient measure, in the
case that we are not capable of completely handling the non-local
effect in the calculation of dark energy perturbations in these
models. But, on the other hand, since the non-local effect is not a
dominant factor, it is believed that the fluid approximation in the
holographic-type dark energy models is reasonable.

\section{Global fit results}\label{sec3}

In our global fit analysis, the parameter $n$ of the NADE model is
treated as a derived parameter, whose value can be obtained by the
parameters $\Omega_{b0}$, $\Omega_{dm0}$ and $\Omega_{k0}$ via the
Newton iteration algorithm, as mentioned in Sec.~\ref{sec2}. The
``CosmoMC'' package uses the physical densities of baryon
$\omega_{b}\equiv\Omega_{b0}h^{2}$ and cold dark matter
$\omega_{dm}\equiv\Omega_{dm0}h^{2}$ instead of $\Omega_{b0}$ and
$\Omega_{dm0}$. Thus, our most general parameter space vector is:
\be \label{parameter} {\bf P} \equiv (\omega_{b}, \omega_{dm},
\Theta, \tau, \Omega_{k0}, n_{s}, A_{s}), \ee where $\Theta$ is the
ratio (multiplied by 100) of the sound horizon to the angular
diameter distance at decoupling, $\tau$ is the optical depth to
re-ionization, $A_s$ and $n_s$ are the amplitude and the spectral
index of the primordial scalar perturbation power spectrum. For the
pivot scale, we set $k_{s0}=0.002$Mpc$^{-1}$ to be consistent with
the WMAP team~\cite{WMAP7}. Note that we have assumed purely
adiabatic initial conditions.

The sound speed of dark energy, $c_s^2\equiv\delta
p_{de}/\delta\rho_{de}$, is usually treated as a phenomenological
parameter for a fluid model of dark energy. It must be real and
non-negative to avoid unphysical instabilities. In fact, if we treat
the dark energy strictly as an adiabatic fluid, then the sound speed
$c_s$ would be imaginary (in this case, the physical sound speed is
equal to the adiabatic sound speed, $c_s^2=c_a^2<0$), leading to
instabilities in dark energy. In order to fix this problem, it is
necessary to assume that dark energy is a non-adiabatic fluid and
impose $c_s^2>0$ by hand~\cite{Gordon:2004ez}. Our analysis is
insensitive to the value of $c_s$. As long as $c_s$ is close to 1,
the dark energy does not cluster significantly on sub-Hubble scales.
Therefore, in our analysis, we set $c_s$ to be 1; the adiabatic
sound speed can be imaginary, $c_a^2\equiv dp_{de}/d\rho_{de}<0$.
This is what is done in the CAMB and CMBFAST codes. Of course, one
can also take $c_s$ as a parameter, but the fit results would not be
affected by this treatment~\cite{Li:2010ac}.

For the data, besides the CMB data (including the WMAP 7-yr
temperature and polarization power spectra~\cite{WMAP7}), the main
other astrophysical results that we shall use in this paper for the
joint cosmological analysis are the following distance-scale
indicators:
\begin{itemize}
\item
The Union2.1 sample of 580 SNIa with systematic errors considered~\cite{Suzuki:2011hu}.
\item
The BAO data, including the measurements of $r_s(z_d)/D_V(0.2)$ and
$r_s(z_d)/D_V(0.35)$ from the SDSS DR7~\cite{Percival:2009xn} and
the measurements of $A$ parameter at $z=0.44,\ 0.6,\ 0.73$ from the
WiggleZ Dark Energy Survey~\cite{Blake:2011en}.
\item
The Hubble constant measurement $H_0=73.8\pm 2.4 {\rm km/s/Mpc}$
from the WFC3 on the HST~\cite{Riess:2011yx}.
\end{itemize}

\begin{table}\caption{Global fit results of the new agegraphic dark energy model.}\label{table1}
\begin{center}
\begin{tabular}{cc cc cc cc}
\hline\hline Class & Parameter & & Best fit with errors  &\\
\hline
Primary &$\Omega_{dm0}h^2$ & &$0.113^{+0.006+0.011+0.015}_{-0.005-0.009-0.014}$&\\
&$100\Omega_{b0}h^2$&&$2.262^{+0.064+0.114+0.158}_{-0.045-0.103-0.156}$&\\
&$\Omega_{k0}$      &&$0.020^{+0.006+0.010+0.016}_{-0.005-0.012-0.016}$&\\
&$\tau$             &&$0.085^{+0.015+0.032+0.049}_{-0.011-0.023-0.036}$&\\
&$\Theta$           &&$1.040^{+0.002+0.005+0.008}_{-0.002-0.005-0.007}$&\\
&$n_s$              &&$0.970^{+0.013+0.025+0.036}_{-0.011-0.023-0.037}$&\\
&$\log[10^{10}A_s]$ &&$3.181^{+0.040+0.080+0.122}_{-0.033-0.075-0.113}$&\\
\hline
Derived &$n$        &&$2.673^{+0.053+0.127+0.199}_{-0.077-0.151-0.222}$&\\
&$H_0$ (km/s/Mpc)   &&$69.2^{+1.2+2.6+3.9}_{-1.4-2.9-4.2}$&\\
&$\Omega_{de0}$      &&$0.697^{+0.011+0.025+0.037}_{-0.015-0.031-0.046}$&\\
&$\Omega_{m0}$      &&$0.284^{+0.014+0.029+0.043}_{-0.010-0.022-0.033}$&\\
&Age (Gyr)          &&$13.125^{+0.202+0.457+0.657}_{-0.220-0.414-0.580}$&\\
&$z_{re}$           &&$10.531^{+1.238+2.485+3.696}_{-0.884-2.058-3.286}$&\\
\hline
\end{tabular}
\end{center}
\end{table}

\begin{figure*}[htbp]
\centering \noindent
\includegraphics[width=17cm]{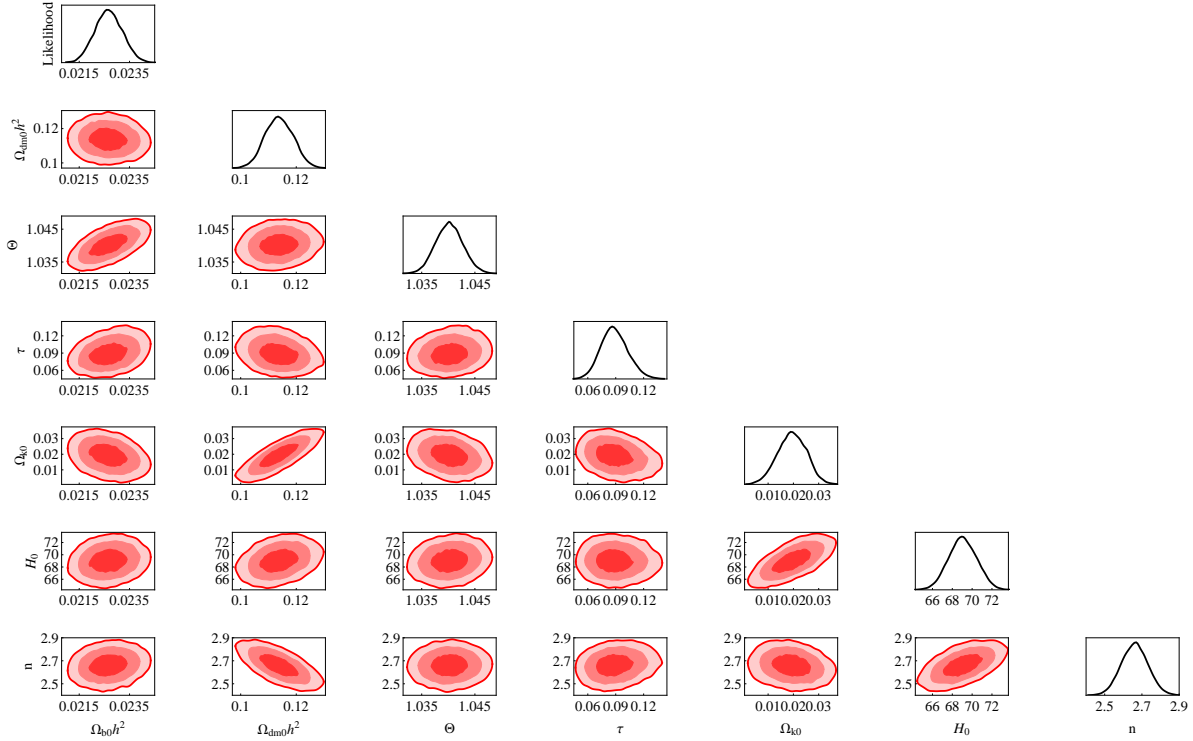}
\caption{\label{fig:NADE}The 1D marginalized distributions of
individual parameters and 2D contours at the 1--3$\sigma$ confidence
levels from the global fit using the CMB+BAO+SNIa+$H_0$ data.}
\end{figure*}

\begin{figure}[htbp]
\centering \noindent
\includegraphics[width=7cm]{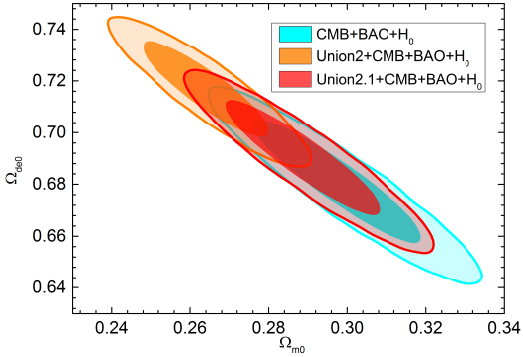}
\caption{\label{fig:diffSN}Parameter spaces in the
$\Omega_{de0}$--$\Omega_{m0}$ plane constrained by different joint
data sets. The cyan, orange and red regions denote the constraint
results from the joint CMB, BAO and $H_0$ observations without SNIa
data, with Union2 sample of SNIa data and with Union2.1 sample of
SNIa data, respectively. Note that the Union2.1 sample is used with
the systematic errors while the Union2 sample isn't, so it is no
wonder why the Union2 sample slightly constrains the parameters more
tightly than the Union2.1 sample.}
\end{figure}

\begin{figure}[htbp]
\centering \noindent
\includegraphics[width=7cm]{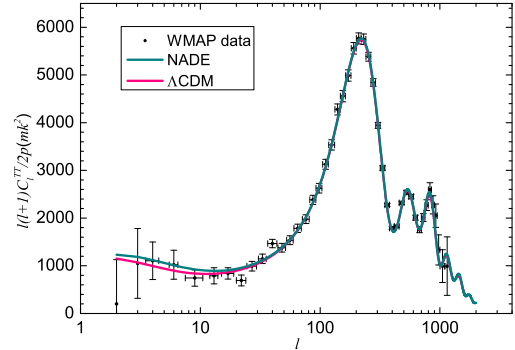}
\caption{\label{fig:power}The CMB $C^{TT}_l$ power spectra for the
NADE and $\Lambda$CDM models with the best-fit parameter values. The
black dots with error bars denote the observational data from the
WMAP 7-yr observations.}
\end{figure}

Our global fit results are summarized in Table \ref{table1} and
Fig.~\ref{fig:NADE}. We notice two remarkable results: (i) the value
of $\Omega_{k0}$ is greater than $0$ at least at the 3$\sigma$
confidence levels (CLs), which implies that the NADE model
distinctly favors an open universe; (ii) the best-fit value of $n$
is evidently less than those obtained in the previous work.

The result of the spatial curvature we obtained is still at the
level of $10^{-2}$, consistent with the other results about dark
energy in the literature. However, it should be pointed out that our
work is the first to show that the NADE model distinctly favors an
open universe under current data.

The fit result of $n$ is
$n=2.673^{+0.053+0.127+0.199}_{-0.077-0.151-0.222}$ at the
1--3$\sigma$ CLs. We find that the best-fit value of $n$ is
significantly different from the results obtained in the previous
work~\cite{Li:2012xm,Wei:2007xu,Li:2009bn,Li:2009jx,Li:2010ak,Wei:2010wu}.
For example, Refs.~\cite{Li:2012xm}, \cite{Li:2009bn},
\cite{Wei:2010wu} and \cite{Li:2010ak} gave $n=2.810$, 2.807, 2.887,
and 2.886, respectively. What can this change tell us? Next, we
shall explore what the factor is that is making this change.

To do so, and without loss of generality, we take the most recent
work in Ref.~\cite{Li:2012xm} as an example to make a comparison. We
list here four different points between them: (1) the spatial
curvature is considered in this work while it isn't considered in
Ref.~\cite{Li:2012xm}; (2) the Hubble constant $H_0$ measurement and
the BAO measurement of $A$ parameter from the WiggleZ Dark Energy
Survey are used in this work while they are not used in
Ref.~\cite{Li:2012xm}; (3) the SNIa data set used in this work is
Union2.1 sample while it is Union2 sample in Ref.~\cite{Li:2012xm};
(4) this paper performs a global fit where the full CMB information
from WMAP-7yr observations is used while the CMB information used in
Ref.~\cite{Li:2012xm} comes only from 7-yr WMAP ``distance priors''.
After testing these four items one by one by varying one condition
and setting the others the same, we find that the change is mainly
due to the different SNIa data used.

To see clearly, we plot the test results of different SNIa data used
for the NADE model in Fig.~\ref{fig:diffSN}. The cyan, orange and
red regions denote the parameter spaces constrained by using the
joint CMB, BAO and $H_0$ observations without SNIa data, with Union2
sample of SNIa data, and with Union2.1 sample of SNIa data,
respectively. Note that the CMB data used here are the 7-yr WMAP
``distance priors'' for simplicity. Comparing the cyan region
(without SNIa) with the orange region (with Union2), one can easily
find that their overlapping region is very small, which implies that
there is a tension between the Union2 sample of SNIa and the joint
CMB, BAO and $H_0$ observations. However, the tension between SNIa
and CMB+BAO+$H_0$ will be greatly relieved once the Union2 sample is
replaced by the Union2.1 sample of SNIa, since the overlap between
cyan and red regions is very large. Thus, we can conclude that the
new constraint result of the NADE model obtained in this work is
more reasonable than before.

In addition, by using the global fit results obtained in this work
(best-fit values), we plot the CMB power spectrum for the NADE model
in Fig.~\ref{fig:power}. To make a comparison, we also plot the
corresponding curve of the $\Lambda$CDM model with the best-fit
parameters given by the same data sets. The 7-yr WMAP observational
data points with uncertainties are also marked in the figure. We can
clearly see that the NADE model produces a different feature
compared with the $\Lambda$CDM model in the low $l$ regions.
However, since the low $l$ data are rare and rough, the two models
cannot be discriminated by the late-time ISW effects by far.

\section{Concluding remarks}\label{sec4}

In this work, our main task is to perform a global fit analysis on
the NADE model in a non-flat universe by using a MCMC method with
the joint data of the full CMB spectra, BAO, SNIa and $H_0$
observations. We use the initial condition,
$\Omega_{de}(z_{ini})=\frac{n^2(1+z_{ini})^{-2}}{4}\left(1+\sqrt{F(z_{ini})}\right)^2$
at $z_{ini}=2000$, to solve the differential equation of
$\Omega_{de}$, which can reduce one free parameter for the NADE
model. In order to easily modify the ``CosmoMC'' package, we set
$\Omega_{dm0}$, $\Omega_{b0}$, and $\Omega_{k0}$ as free parameters
and $n$ as a derived parameter, and use the Newton iteration
algorithm to obtain the true value of $n$.

Our global fit results show that the NADE model distinctly favors an
open universe, since the value of $\Omega_{k0}$ is greater than $0$
at the 3$\sigma$ CLs. Besides, we find that the value of the key
parameter of the NADE model
$n=2.673^{+0.053+0.127+0.199}_{-0.077-0.151-0.222}$ significantly
different from those obtained in the previous work. We find that the
main reason leading to such a change comes from the different SNIa
samples used. Namely, the Union2.1 sample and the Union2 sample give
fairly different constraint results for the NADE model. Our further
test indicates that there is a distinct tension between the Union2
sample of SNIa and other observations, and the tension will be
removed once the Union2 sample is replaced by the Union2.1 sample.
Thus, we conclude that the new constraint result of the NADE model
obtained in this paper is more reasonable than the previous ones.

We also plot the CMB power spectra for the NADE and $\Lambda$CDM
models by using the best-fit results of the global fits. It is shown
that the NADE model and the $\Lambda$CDM model produce slightly
different features in the low $l$ regions. Since the low $l$ data
are rather rare and rough, the two models cannot be discriminated by
the late-time ISW effects yet. We expect that the future highly
accurate data can do this job.

If dark energy is truly dynamical, as hinted by recent
work~\cite{Zhao:2012aw}, the possible non-gravitational interaction
between dark energy and dark matter may deserve deeper
investigations. In this work the interaction between dark energy and
dark matter is not considered. If such an interaction exists in the
NADE model, the calculations of the perturbations of dark energy and
dark matter will become more complicated. The global fit analysis on
the interacting agegraphic dark energy model is fairly attractive.
In addition, the comparison of the various holographic models of
dark energy under a uniform global fit is also necessary. We leave
these subjects in the future work.

\begin{acknowledgments}
This work was supported by the National Science Foundation of China
under Grant Nos.~10705041, 10975032, 11047112 and 11175042, and by
the National Ministry of Education of China under Grant
Nos.~NCET-09-0276, N100505001, N090305003, and N110405011.
\end{acknowledgments}

\end{document}